\documentclass[a4paper]{article}

\usepackage{INTERSPEECH2022}
\usepackage{multirow}
\usepackage{hyperref}
\usepackage{svg}
\usepackage{color}
\usepackage{threeparttable}
\usepackage{colortbl}
\usepackage{url}
\usepackage{graphicx}
\usepackage{adjustbox}
\usepackage{amssymb}      
\title{Text-Aware End-to-end Mispronunciation Detection and Diagnosis}
\name{Linkai Peng$^1$, Yingming Gao$^1$, Binghuai Lin$^2$, Dengfeng Ke$^1$, Yanlu Xie$^1$, Jinsong Zhang$^1$}
\address{
  $^1$School of Information Sciences, Beijing Language and Culture University, Beijing, China\\
  $^2$Smart Platform Product Department,Tencent Technology Co., Ltd, Beijing, China
  }
\email{penglinkai96@gmail.com, yingming.gao@mailbox.tu-dresden.de, binghuailin@tencent.com,\\ \{dengfeng.ke, xieyanlu, jinsong.zhang\}@blcu.edu.cn}

\begin{document}

\maketitle
\begin{abstract}
Mispronunciation detection and diagnosis (MDD) technology is a key component of computer-assisted pronunciation training system (CAPT). In the field of assessing the pronunciation quality of constrained speech, the given transcriptions can play the role of a teacher. Conventional methods have fully utilized the prior texts for the model construction or improving the system performance, e.g. forced-alignment and extended recognition networks. Recently, some end-to-end based methods attempt to incorporate the prior texts into model training and preliminarily show the effectiveness. However, previous studies mostly consider applying raw attention mechanism to fuse audio representations with text representations, without taking possible text-pronunciation mismatch into account. In this paper, we present a gating strategy that assigns more importance to the relevant audio features while suppressing irrelevant text information. Moreover, given the transcriptions, we design an extra contrastive loss to reduce the gap between the learning objective of phoneme recognition and MDD. We conducted experiments using two publicly available datasets (TIMIT and L2-Arctic) and our best model improved the F1 score from 57.51\% to 61.75\% compared to the baselines. Besides, we provide a detailed analysis to shed light on the effectiveness of gating mechanism and contrastive learning on MDD\footnote{\url{https://github.com/vocaliodmiku/wav2vec2mdd-Text}}.
% misleading the model towards sub-optimization. 
\end{abstract}

\section{Introduction}
Computer-Assisted Pronunciation Training (CAPT) system can meet people's needs for language learning in fragmented time with flexible devices. Mispronunciation detection and diagnosis system (MDD) is an indispensable component of the CAPT system. Similar to the role of teachers in oral practice lessons, MDD can provide instant feedback about pronunciation problem for users to improve their speaking skills. Considering the rapidly increasing number of language learners, a high-performance MDD is needed to assure the precise diagnosis of pronunciation errors at the phonetic and prosodic levels. Here, we focus on phonetic mispronunciation in second-language learning.

Here, we consider assessing the pronunciation quality of constrained speech, that is, the text uttered by speakers is known to the system. Popular pronunciation error detection framework can be roughly divided into two categories, both of which have fully made use of the transcriptions. The first category is based on confidence measures which are mainly obtained from automatic speech recognition (ASR). Whether the pronunciation is correct or not is decided by calculating the confidence score of the frame/phoneme level with the help of forced alignment \cite{witt2000phone,hu2013new,zheng2007generalized}, which is a technique to align acoustic frames with given texts. The second category is based on extended search lattice and one of the most popular approaches is extended recognition network (ERN) \cite{harrison2009implementation}. ERN analyzes the text first and then incorporates a finite number of phonetic error patterns into the decoding network based on handcrafted or data-driven rules. 
%However, some limitations exist in these methods: confidence measures based approaches lacks the ability to provide specific diagnostic information; ERN cannot guarantee that all mispronunciations are covered and unseen mispronunciations will lead bad performance; Multistage systems have complex structures and the building process is laborious; 
%Recently, inspired by deep neural network (DNN) ASR framework, the CTC-based End-to-End (E2E) method was applied to MDD and achieved promising performance. The end-to-end modeling approach avoids the complicated modeling process and hand-designed dictionaries. However, this kind of method usually need a large amount of labeled data to make all the weights in network fully trained while MDD is a data-scarce task. Large-scale non-native speech data is difficult to collect and the corresponding annotations rely on the support of experienced experts. Moreover, legal risk to individual privacy during audio recordings should also be taken into consideration.

Recent studies have proposed various network architectures to improve the MDD performance \cite{leung2019cnn,Yan2020,zheng2021cca,feng2020sed,fu2021full,jiang2021towards}. Similar to conventional methods, a line of studies attempt to leverage the prior linguistic information to provide guidance extracted from the given transcriptions \cite{zheng2021cca,feng2020sed,fu2021full,jiang2021towards}. \cite{feng2020sed} feeds phoneme sequences into a sentence encoder and then combines with audio features via attention. Frame-level cross-entropy loss is calculated with the help of manually labeled phoneme boundary. \cite{fu2021full} designs multiple data augment techniques based on the given transcriptions to alleviate the data sparsity problem. Despite the effective network design, most previous studies directly incorporate textual features into speech representation via a naive attention mechanism. We contend that textual features contribute very differently when they are assigned to attend different acoustic features. For correct pronunciation, transcription can guide the model step towards text-audio joint representation for better inference. However, it is difficult to align prior phonemes with acoustic features when mispronunciation occurs and hence limits the potential performance improvement.

End-to-end MDD shows its success in modeling simplicity and performance improvement. The main idea is to train a phoneme recognition model on L1/L2/L1-L2 hybrid datasets, and then perform MDD by comparing the reference and the inference. However, most previous models are optimized with a sole phoneme recognition objective directly or implicitly constrained by extra error states towards the correct diagnosis. Such a single recognition loss tries to predict each phoneme equally. To some extent, we hope the system can report more mispronunciations with little/no sacrifice of performance on canonicals. Some works utilize an extra error-state-related loss function to carry out MDD implicitly \cite{zheng2021cca}. Due to the gap between the learning objective of phoneme recognition and MDD, previous methods fail to focus on mispronunciations explicitly and thus being less effective in detection and diagnosis.

In this study, we propose a \emph{Text-Aware end-to-end} model for MDD, which incorporates the prior text modality to learn a good joint representation of acoustic units. We leverage an effective Text-Audio gate control module to effectively fuse prior transcriptions. It can enforce the model to align textual information to the most related acoustic regions while ignoring irrelevant parts automatically. To further unleash the power of prior texts, we refine the loss to bridge the learning objective gap between phoneme recognition and MDD by explicitly discriminating the probability of reference and annotation sequences. We experiment on the L2-Arctic dataset. Results confirm our main hypothesis that modeling text with gate control and explicitly distinguishing the reference and annotation benefit performance.

\section{Text-Aware E2E MDD}
\begin{figure}[h]
\centering
  \includegraphics[width=1\linewidth]{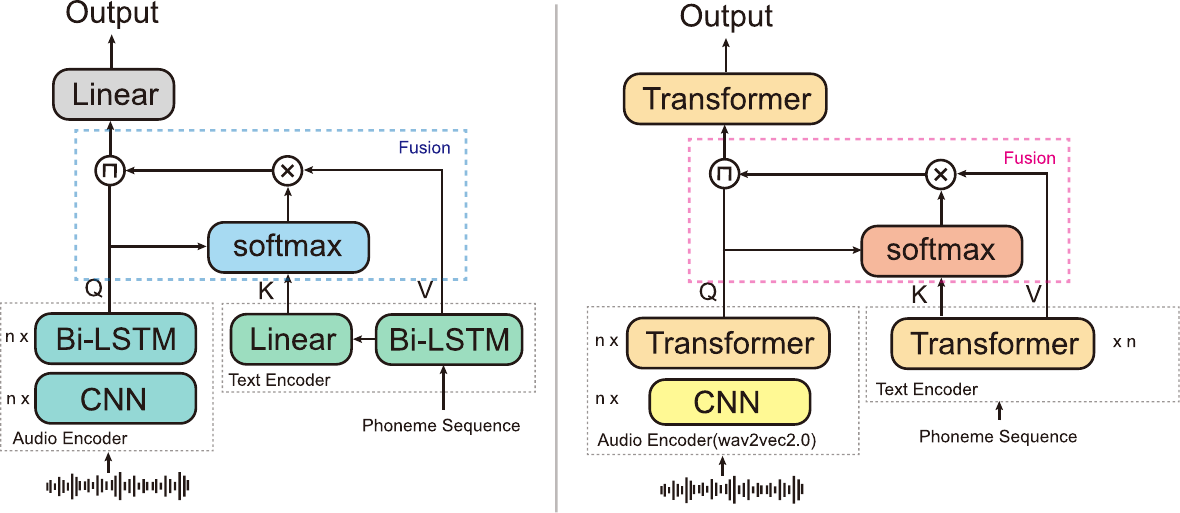}
  \caption{An overview of baselines: \textbf{(left)} Baseline comes from \cite{fu2021full}. The framework takes phoneme sequences and the fbank feature as inputs and improves the aligned representation learning by performing attention for feature aggregation. \textbf{(right)} Our baseline model. We replace the audio encoder and text encoder with more powerful wav2vec2.0 series and Transformer while keeping the same fusion strategy. Note that only the audio branch is pretrained.}
  \label{fig:intro}
\end{figure}

We first briefly introduce the model structure we used for exploring information control. Then we explain the notion of contrastive learning object which will be performed for E2E MDD. Due to space limitations, we scatter the proposed architecture into Figure~\ref{fig:intro} and Figure~\ref{fig:gate}. The pre-trained model we used comes from fairseq toolkit \cite{ott2019fairseq}.

\subsection{Audio encoder}
In our previous work \cite{peng21e_interspeech} and work \cite{xu2021explore} from another group, pretrained acoustic model  have achieved great success on MDD. Here we inherit the pretrained model wav2vec2.0 \cite{baevski2020wav2vec} as audio encoder. It consists of a CNN-based encoder network, a transformer-based context network and a vector quantization module. We omit quantization module because it is out of the scope of this paper. The encoder network $\boldsymbol{f}:\mathcal{X}\mapsto\mathcal{Z}$ encodes the raw audio sample point $x_i \in \boldsymbol{X}$ into latent speech representation $[\mathbf{z}_1,\mathbf{z}_2,...,\mathbf{z}_T]$. Combining multiple layer normalization and GELU activation layers, the convolutional module compresses about 25 ms of 16 kHZ audio every 20 ms. Then context representations $\mathbf{H}^{\mathrm{audio}}=\boldsymbol{g}([\mathbf{z}_1,\mathbf{z}_2,...,\mathbf{z}_T])$ are obtained by a context network $\boldsymbol{g}:\mathcal{Z} \mapsto \mathcal{H}$ which  scans over the entire latent speech representations.

\subsection{Text encoder}

Although we have included such a powerful acoustic model into our MDD system, there is still possibly more room for improvement by combining with reference texts. A common approach is to convert the reference text into phonemes and transform each phoneme into an N-dimensional linguistic feature vector via a parameterized lookup table. We construct the text encoder with a network of two Transformer layers. Given a canonical phoneme sequence with length $N$, the reference text representation $\mathbf{H}^{\mathrm{text}}=[\mathbf{h}^{\mathrm{text}}_1,\mathbf{h}^{\mathrm{text}}_2,...,\mathbf{h}^{\mathrm{text}}_N]$ can be derived by the text encoder.

\subsection{Textual modulation gate}\label{sec:gate}
\begin{figure}[h]
\centering
  \includegraphics[width=1.05\linewidth]{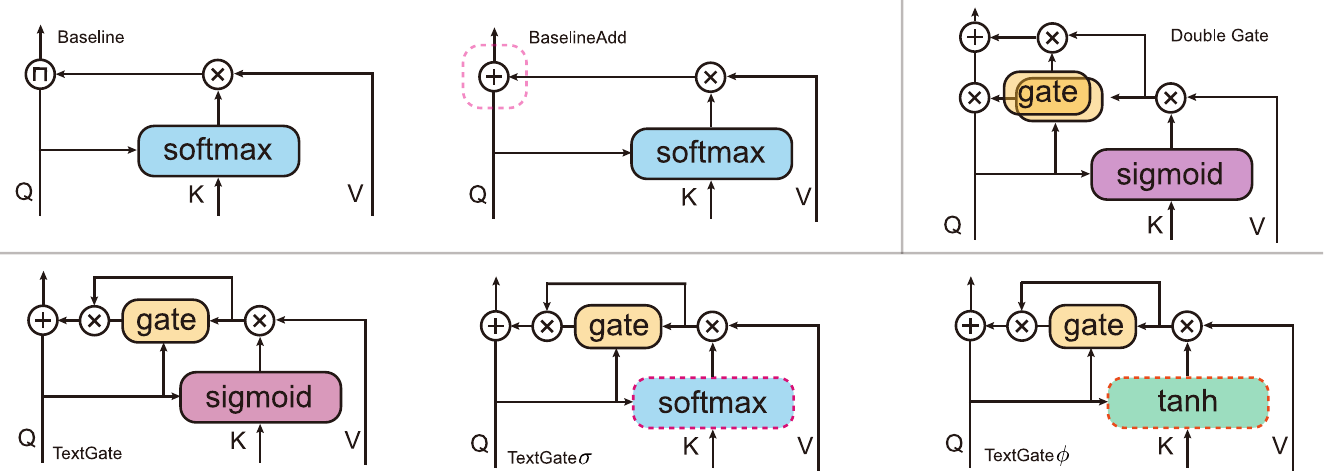}
  \caption{Variants of textual-gate modulating: \textbf{(Baseline)} is identical with Figure~\ref{fig:intro}. \textbf{(BaselineAdd)} uses another popular feature fusion operation “add” instead. \textbf{(Double Gate)} not only performs control on textual branch, but also monitors acoustic branch with another parameter setting. \textbf{(TextGate)} is presented in Section sec~\ref{sec:gate}. \textbf{(TextGate$\sigma$)} and \textbf{(TextGate$\phi$)} attempt to look into different activation functions.}
  
  \label{fig:gate}
\end{figure}

In accordance with the postulates given in the introduction, we design a \emph{Textual Modulation Gate} based on attention fusion. Compared with the annotation transcription, some phonemes in reference text are replaced with the corresponding prompts which reflect the actual acoustic parameters. The ``polluted" reference text is thus not paired with associated audio features. On the textual side, we run an information monitor to filter out texts whose prior knowledge is strong enough to deteriorate the performance. For $\mathbf{H}^{\mathrm{text}}$ and $\mathbf{H}^{\mathrm{audio}}$, we have:

\begin{gather}
\mathbf{\alpha}_{t,n} = \mathrm{sigmoid}(\mathrm{score}(\mathbf{h}^{\mathrm{text}}_n, \mathbf{h}^{\mathrm{audio}}_t)) \\
\mathrm{score}(\mathbf{h}^{\mathrm{text}}_n, \mathbf{h}^{\mathrm{audio}}_t) = \mathbf{h}^{\mathrm{audio}}_n (\mathbf{h}^{\mathrm{text}}_t)^{\mathbf{T}} \\ % audio [100,768] text [50,768]  [100,50]
\mathbf{c}_t = \sum^N_{n=1} \mathbf{\alpha}_{t,n}\mathbf{h}^{\mathrm{text}}_n \\ % ct [100,768]
\mathbf{g} = \mathrm{sigmoid}(W\cdot \mathbf{h}^{\mathrm{audio}}_t  + U \cdot \mathbf{c}_t + b) \\ 
\mathbf{y}_t = \mathbf{h}^{\mathrm{audio}}_t + \mathbf{g} \odot \mathbf{c}_t
\end{gather}

\noindent where $\odot$ is element-wise product. We compute attention weight between frame $\mathbf{h}^{\mathrm{text}}_n \in \mathbf{H}^{\mathrm{text}}$ and $\mathbf{h}^{\mathrm{audio}}_t \in \mathbf{H}^{\mathrm{audio}}$ which is used for re-weighting the textual representation. Then we choose the implementation of linear projection, summation, and sigmoid activation sequentially to generate the textual gate before feeding them into the Transformer layer for CTC prediction. We refer to the formula above as \emph{TextGate}. Furthermore, we further explore variants of gate modulating (Figure~\ref{fig:gate}) and conduct experiments to evaluate them. 

\subsection{Contrastive learning}
Based on our experimental results, we found that better phoneme recognition model implementations can not always report better results in the context of MDD. In Figure~\ref{fig:baseline}, all the results in terms of phone error rate and F1 score are leaked in advance. As the phone error rate decreases, the F1 score performance trend is hard to conclude. The failure is due to the mismatch of learning objectives between MDD and phoneme recognition. Phoneme recognition aims to infer phonemes from the annotation correctly as much as possible, irrespective of whether we should pay more attention to mispronunciations. In this situation, performance improvement in recognition can be achieved by detecting more canonicals in proportion. With the given prior texts, we propose an objective base on contrastive learning to bridge the gap. Contrastive learning, a kind of technique that maximizes the intra-class similarity and minimizes the inter-class similarity has been used extensively over the years in various applications \cite{jaiswal2020survey,le2020contrastive}. In the context of MDD, we can anchor the transcription in order to generate the dissimilarity/similarity. 

\begin{figure}[h]
\centering
  \includegraphics[width=1\linewidth]{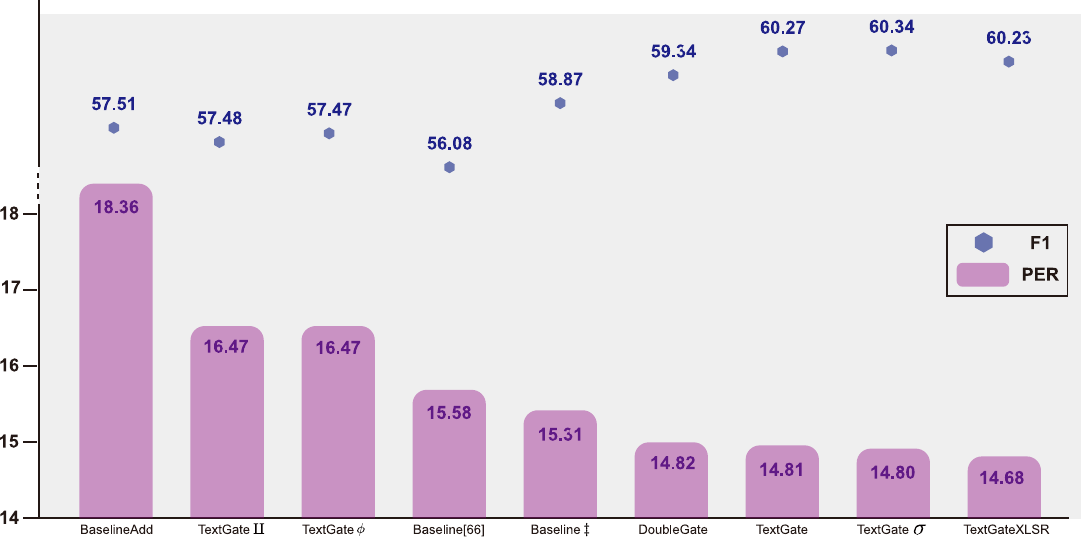}
  \caption{Performance comparison among our experiments. We present \textbf{P}hone \textbf{E}rror \textbf{R}ate and \textbf{F1} score here. }
  \label{fig:baseline}
\end{figure}

While we cannot directly construct negative pairs and positive pairs as usual to define the similarity, we introduce a supervised contrastive loss derived from CTC \cite{graves2006connectionist}. Addressing the variable length (T) input frames, $\boldsymbol{X}=[\boldsymbol{x}_1,\boldsymbol{x}_2,...,\boldsymbol{x}_T]$, U length associated reference characters, $\boldsymbol{L}^e$, and U length associated reference characters, $\boldsymbol{L}$, conditionally independent probability of label sequence:
\begin{gather}\label{eqn:con01}
\begin{aligned}
p(\boldsymbol{\pi}|\boldsymbol{X}) &= \prod^T_{t=1} y_{\pi_t}^t = y_{\pi_1}^1 y_{\pi_2}^2...y_{\pi_t}^t...y_{\pi_T}^T, \forall \boldsymbol{\pi} \in \Phi(L). \\
p(\boldsymbol{\pi}^e|\boldsymbol{X}) &= \prod^T_{t=1} y_{\pi^e_t}^t = y_{\pi^e_1}^1 y_{\pi^e_2}^2...y_{\pi^e_t}^t...y_{\pi^e_T}^T, \forall \boldsymbol{\pi} \in \Phi(L^e).\nonumber
\end{aligned}
\end{gather}
\noindent where $y_{\pi_t}^t$ denotes the softmax output of label $\pi_t$ at time t, $\Phi(\cdot)$ is a map function which can generate all possible intermediate label representations from unmodified label sequence. A modified label sequence $\boldsymbol{\pi}$ is made by inserting the blank symbols between each label including the beginning and the end (i.e., $\boldsymbol{L}=[c,a,t],\boldsymbol{\pi}=[-,c,-,a,-,t,-]$). Suppose there is only one substitution mispronunciation occurred at position t, for each possible $\pi^e$, we can obtain a paired $\pi$ from $\Phi(L)$ and therefore, for paired $\pi$-$\pi^e$, $y_{\pi_k}^k=y_{\pi^e_k}^k$, $\forall k \neq t$. Then we can define the dissimilarity for modified annotation and sequence.
\begin{equation}\label{eqn:con02}
D^{\pi,\pi^e,X}_{\mathrm{contrast}} = \ln p(\pi^e|X) - \ln p(\pi|X) = \ln y_{\pi^e_t}^t - \ln y_{\pi_t}^t.
\end{equation}

We incorporate margin into the dissimilarity and sum up all possible negative pairs. Then our contrastive loss can be expressed as:
\begin{equation}\label{eqn:asr07}
\mathcal{L}_{contrast}^{(L,L^e,\boldsymbol{X})}  \triangleq \sum_{\substack{\pi,\pi^e \in \\ B^{-1}(L,L^e)}} \max(\ln p(\pi^e|\boldsymbol{X}) - \ln p(\pi|\boldsymbol{X}) + m, 0).\nonumber
\end{equation}
In order to train the network, we use besides the contrastive loss $\mathcal{L}_{contrast}$ one additional loss functions:
\begin{equation}\label{eqn:asr08}
\mathcal{L} = \mathcal{L}_{CTC} + \mathcal{L}_{contrast}.
\end{equation}

\noindent where $\mathcal{L}_{CTC}$ is a loss for phoneme sequence recognition\footnote{CTC computes the probability of all possible intermediate sequence via dynamic programming. Similarly, we implemented the loss as followed  approximately. $m$ is set to 16 empirically :
\begin{equation}\label{eqn:asr07}
\mathcal{L}^{(L,L^e,\boldsymbol{X})}_{\mathrm{contrast}} = \max (\ln p(L^e|\boldsymbol{X}) - \ln p(L|\boldsymbol{X}) + m,0).
\end{equation}}. In practice, contrastive learning can be realized easily on an attention-based encoder-decoder model for the convenience of accessing phoneme-level likelihoods. Based on our experimental results, we find that the encoder-decoder structure cannot achieve better results and conclude that the data used in this paper is too sparse for the decoder to generalize well. An alternative approach to reducing the gap is to directly optimize the F1 score metric with the reinforcement learning technique \cite{kaelbling1996reinforcement}.

\begin{table*}

\centering
\begin{adjustbox}{max width=0.85\textwidth}
\begin{threeparttable}
    \caption{Comprehensive performance comparison. }
    \begin{tabular}{lcccccc}
    
\toprule
\toprule
    \multirow{3}{*}{Models} & 
    \multicolumn{2}{c}{Canonicals~} & \multicolumn{3}{c}{Mispronunciations~} & \multirow{3}{*}{F1} 
    \\ 
    \cline{2-6} & 
    \multirow{2}{*}{True Accept} & 
    \multirow{2}{*}{False Rejection} & 
    \multirow{2}{*}{False Accept} & 
    \multicolumn{2}{c}{True Rejection}               
    \\ 
    \cline{5-6}
    &&&&
    Corroct Diag. & 
    Diag. Error
    \\ 
    \toprule
    Baseline$\dagger$ & 
    92.65\% (23825) & 
    7.35\% (1889) &  
    43.88\% (1883) &  
    74.96\% (1805) &  
    25.04\% (603) &
    56.08\% % 15.58
    \\
    
    \rowcolor{gray!9}
    Baseline$\ddagger$ & 
    93.89\% (24172) & 
    6.11\% (1574) &  
    42.87\% (1826) &  
    68.52\% (1667) &  
    31.48\% (766) &
    58.87\% % 15.31 baseline0 cat 
    \\
    \rowcolor{gray!9}
    BaselineAdd& 
    94.15\% (24239) & 
    5.85\% (1507) &  
    45.36\% (1932) &  
    63.21\% (1471) &  
    36.79\% (856) &
    57.51\% % 18.36 baseline add
    \\
    \hline
    Double Gate & 
    94.59\% (24352) & 
    5.41\% (1394) &  
    44.00\% (1874) &  
    68.68\% (1638) &  
    31.32\% (747) &
    59.34\% % 14.82
    \\
    
    \rowcolor{gray!20}
    TextGate & 
    94.50\% (24330) & 
    5.50\% (1416) &  
    42.52\% (1811) &  
    68.26\% (1671) &  
    31.74\% (777) &
    60.27\% % 14.81
    \\
    
    \rowcolor{gray!20}
    TextGate$\sigma$ & 
    94.29\% (24277) & 
    5.71\% (1469) &  
    41.89\% (1784) &  
    69.86\% (1729) &  
    30.14\% (746) &
    60.34\% % 14.80
    \\
    
    \rowcolor{gray!20}
    TextGate$\phi$ & 
    94.53\% (24337) & 
    5.47\% (1409) &  
    46.33\% (1973) &  
    64.22\% (1468) &  
    35.78\% (818) &
    57.48\% % 16.47
    \\
    
    \rowcolor{gray!20}
    TextGate$\sigma R^{\star}$ & 
    95.07\% (24477) & 
    4.93\% (1269) &  
    47.66\% (2030) &  
    63.62\% (1418) &  
    36.38\% (811) &
    57.47\% % 16.473
    \\
    TextGateXLSR & 
    94.94\% (24442) & 
    5.06\% (1304) &  
    43.72\% (1862) &  
    68.79\% (1649) &  
    31.21\% (748) &
    60.23\% % 14.68
    \\
    \hline
    TextGateContrast & 
    93.72\% (24130) & 
    6.28\% (1616) &  
    40.43\% (1722) &  
    69.77\% (1770) &  
    30.23\% (767) &
    60.32\% % 15.68
    \\
    \rowcolor{gray!20}
    TextGateXLSRContrast & 
    93.81\% (24152) & 
    6.19\% (1594) &  
    38.62\% (1645) &  
    71.08\% (1858) &  
    28.92\% (756) &
    61.75\% % 15.06
    \\
    
\bottomrule
\end{tabular}
\begin{tablenotes}
\item[$\dagger$]  The best results reported in \cite{fu2021full}: +VC(10\%).
\item[$\ddagger$] Our baseline, refer to Figure~\ref{fig:intro}.
\item[$\star$] Reversed version of TextGate (maybe we can call it AudioGate), $g$ is used to multiple $h_t^Q$ instead of $c_t$.
\end{tablenotes}
\label{tab:f1}
\end{threeparttable}
\end{adjustbox}
\end{table*}

\section{Experiments}

\subsection{Speech corpora and model architecture}
\noindent \textbf{Datasets.} We use the publicly available datasets TIMIT \cite{garofolo1993darpa} and L2-arctic \cite{zhao2018l2} to conduct our experiments. TIMIT is a native (L1) English corpus containing 6,300 utterances from 630 speakers. We use its original training subset. The L2-arctic corpus\footnote{\url{https://psi.engr.tamu.edu/l2-arctic-corpus/} :L2-ARCTIC-V2.0} is a non-native English speech corpus that is intended for research in voice conversion, accent conversion, and mispronunciation detection. It contains utterances with mispronunciations of 24 (12 males and 12 females) non-native speakers whose L1 languages include Hindi, Korean, Spanish, Arabic, Vietnamese and Chinese. Following prior works \cite{feng2020sed,fu2021full,peng21e_interspeech}, six speakers (NJS, TLV, TNI, TXHC, YKWK, ZHAA) were selected as the test set while the rest were merged to build the training set. We further generated a subset from the training set as the development set by randomly selecting 20\% sentences for each speaker. There was no overlap between training and developing set. For the phone set, we mapped the TIMIT 61-phone to 39-phone according to the mapping table from \cite{lee1989speaker} and combined it into L2-arctic phone set. 

\noindent \textbf{Implementation Details.} For audio encoder, we used two publicly available pre-trained wav2vec2.0 models as our backbones: wav2vec2.0-BASE\footnote{\url{https://dl.fbaipublicfiles.com/fairseq/wav2vec/wav2vec_small.pt}} and wav2vec2.0-XLSR\footnote{\url{https://dl.fbaipublicfiles.com/fairseq/wav2vec/xlsr_53_56k.pt}}. 
All models were trained 142 epochs using the Adam optimizer with an initial learning rate of $5e^{-5}$ on one RTX3060 GPU. The dimension of attention and gating mechanism was set to 768. The audio encoder was frozen in the first 10,000 steps.

\subsection{Performance evaluation}
We followed the evaluation metrics of previous studies \cite{li2016mispronunciation}. For the end-to-end model, the detection of pronunciation errors can be achieved by comparing the prediction sequence and the reference text sequence after alignment. For canonical phones, true accept (TA) means the recognized phone sequence is consistent with the reference text while rejection (FR) means inconsistence. For mispronounced phones, true rejection (TR) indicates the mispronunciation has been detected while false accept (FA) fails to do it. Further, true rejection can be divided into correct diagnosis and diagnosis error. Other metrics like recall (TR/(FA + TR)), precision (TR/(FR + TR)) and the F-1 score (2*((precision*recall)/(precision+recall))) can be calculated based on the accumulated statistics.

\subsection{Experimental results}
We summarized all results on the task of MDD in Table~\ref{tab:f1}.

\noindent \textbf{Evaluating Baselines.}
Under fair architecture and training settings, our Baseline surpasses other baselines. Compared with Baseline from \cite{fu2021full}, audio encoder wav2vec2.0 in our Baseline provides more powerful representations. For a simple fusion operation, concatenation is better than addition. Note that all our systems except BaselineAdd use add implementation. Tuning to concatenation will bring further improvement.

\noindent \textbf{Evaluating Textual Modulation Gate.}
As expected, the proposed approaches \textbf{Double Gate} and \textbf{TextGate} outperform the Baseline/BaselineAdd method by +0.5\%/1.8\% and +1.4\%/2.7\%, respectively. The textual modulation gate successfully plays the role of validating information that comes from texts. Figure~\ref{fig:baseline} shows attention weights output by the \textbf{TextGate} and \textbf{Baseline} model. Since the Textual Modulation Gate can take responsibility for turning on/off textual information flow, attention patterns look neat and natural, while audio-text correlation maps for the model without gate would be chaotic.

\begin{figure}[h]
\centering
  \vspace{-0.2cm}
  \includegraphics[width=1\linewidth]{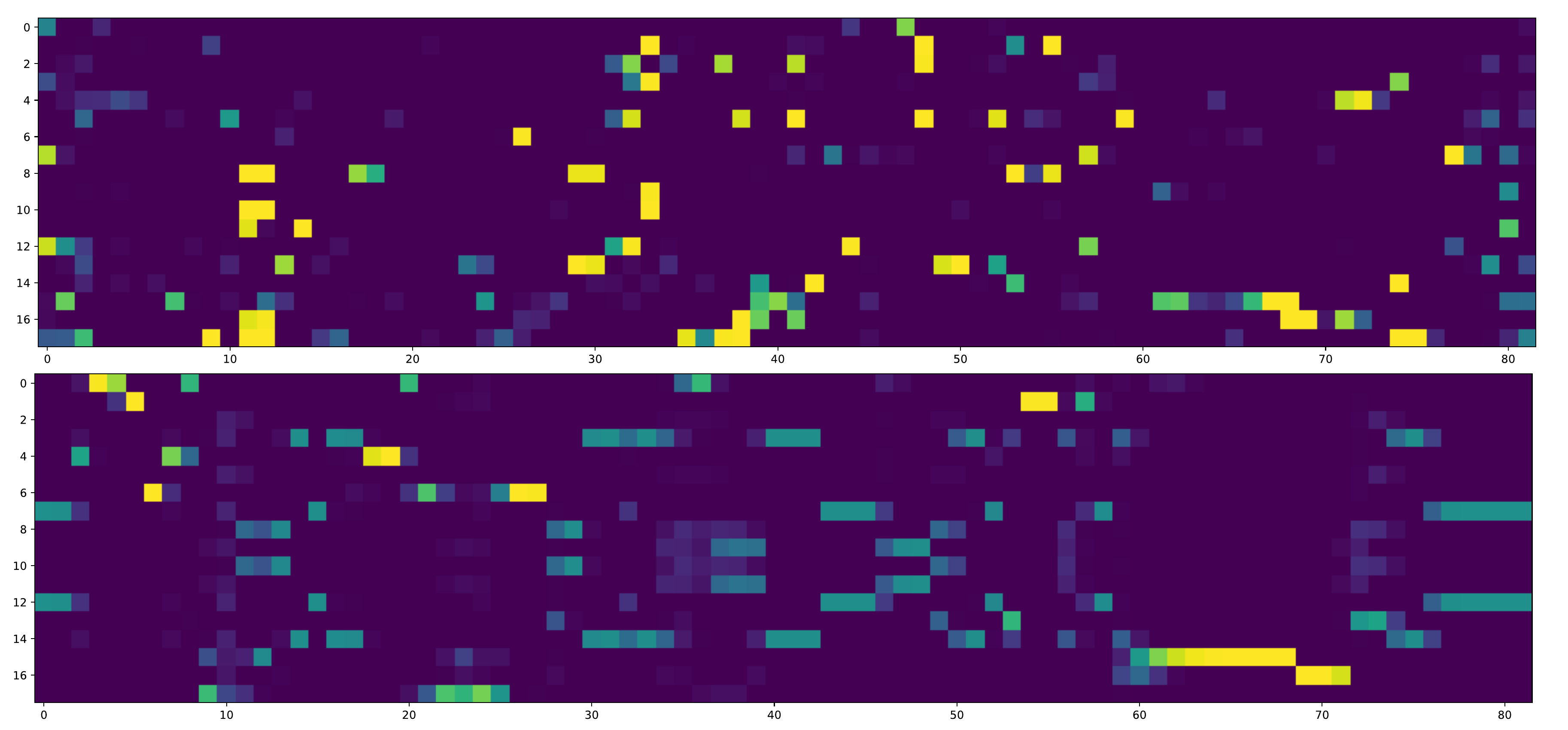}
  \caption{Learned correlation maps of baseline (Up) and TextGate (Down).}
  \label{fig:baseline}
  \vspace{-0.4cm}
\end{figure}

Within the TextGate series, the difference in performance between sigmoid and softmax is small. However, tanh gives a lower performance ever compared to the baseline and we also notice that ``AudioGate", performing control on the audio branch, reports poor performance. These results suggest that even incorporating the extra reference text, the fusion framework needs to be carefully designed. We further tried to utilize a more powerful acoustic model (XLSR) for further improvement and attribute the minor difference to the learning object gap. Relevant result analyses are shown below.

\noindent \textbf{Evaluating Contrastive Learning Object.}
We integrated the proposed \textbf{TextGate} into \textbf{XLSR} model where a contrastive loss was used. As shown in Table~\ref{tab:f1} (bottom), \textbf{TextGateXLSRContrast} obtains a performance gain of 1.5\% F1 score compared model \textbf{TextGateContrast} and achieves the best performance with an F1 score of 61.75\%, suggesting that our methods work practically well with a contrastive loss. Furthermore, \textbf{TextGateXLSRContrast} reports the lowest False Accept number, which corresponds to our discussion — Given the prior text, the model learns more discriminative features about the reference and annotation phonemes which are hard to distinguish, naturally making the prior transcripts more informative.
\section{Conclusion and future work}

In this work, we have presented \emph{Text-Aware end-to-end} model that explicitly incorporates prior transcription in model training and effectively learns a better refined text-audio representation for MDD. With the text-aware module, our best model improves the baseline by +2.8\% in absolute F1 score on L2-arctic dataset. Moreover, we notice the existence of differences between current methods and MDD in optimization objectives. Contrastive-based loss is proposed to bridge the gap and outperforms the baseline methods by +4.24\%. Analyses show that these simple modifications help the mispronunciation-sensitive representation learning among given reading texts and acoustic inputs. In future work, we will investigate  extracting more information from transcriptions, such as transferring phonetic knowledge to constrain the text-audio attention matrix and optimize the learning object toward MDD. 

\section{Acknowledgements}

This study was supported by Advanced Innovation Center for Language Resource and Intelligence (KYR17005), National Social Science Foundation of China (18BYY124), Wutong Innovation Platform of Beijing Language and Culture University (19PT04), the Science Foundation and Special Program for Key Basic Research fund of Beijing Language and Culture University (the Fundamental Research Funds for the Central Universities) (21YJ040004). Jinsong Zhang is the corresponding author.

\bibliographystyle{IEEEtran}

\bibliography{mybib}

% \begin{thebibliography}{9}
% \bibitem[1]{Davis80-COP}
%   S.\ B.\ Davis and P.\ Mermelstein,
%   ``Comparison of parametric representation for monosyllabic word recognition in continuously spoken sentences,''
%   \textit{IEEE Transactions on Acoustics, Speech and Signal Processing}, vol.~28, no.~4, pp.~357--366, 1980.
% \bibitem[2]{Rabiner89-ATO}
%   L.\ R.\ Rabiner,
%   ``A tutorial on hidden Markov models and selected applications in speech recognition,''
%   \textit{Proceedings of the IEEE}, vol.~77, no.~2, pp.~257-286, 1989.
% \bibitem[3]{Hastie09-TEO}
%   T.\ Hastie, R.\ Tibshirani, and J.\ Friedman,
%   \textit{The Elements of Statistical Learning -- Data Mining, Inference, and Prediction}.
%   New York: Springer, 2009.
% \bibitem[4]{YourName17-XXX}
%   F.\ Lastname1, F.\ Lastname2, and F.\ Lastname3,
%   ``Title of your INTERSPEECH 2021 publication,''
%   in \textit{Interspeech 2021 -- 20\textsuperscript{th} Annual Conference of the International Speech Communication Association, September 15-19, Graz, Austria, Proceedings, Proceedings}, 2020, pp.~100--104.
% \end{thebibliography}

\end{document}